\documentclass[12pt]{article}
\usepackage{graphicx}
\usepackage{subfigure}
\usepackage[dvips]{color}
\usepackage{amssymb}
\usepackage{amsmath}

\input epsf.tex

\newcommand{\beq}{\begin{equation}}
\newcommand{\eeq}{\end{equation}}
\newcommand{\be}{\begin{equation}}
\newcommand{\ee}{\end{equation}}
\newcommand{\bea}{\begin{eqnarray}}
\newcommand{\eea}{\end{eqnarray}}

\makeatletter
\renewcommand{\theequation}{\thesection.\arabic{equation}}
\@addtoreset{equation}{section}
\makeatother

\def\href#1#2{#2}
\begin{document}

\baselineskip=15.5pt
\pagestyle{plain}
\setcounter{page}{1}

\begin{titlepage}
\begin{flushleft}
      % \hfill                      {\tt hep-th/1102.****}\\
      % \hfill                       FIT HE - 12-03 \\
       %\hfill                       KYUSHU-HET 135 \\
       %\hfill                       SAGA-HE- \\
\end{flushleft}
%\vspace*{3mm}

\begin{center}
  {\huge Comment on Dark Matter Capture   \\ 
   \vspace*{2mm}
in Neutron Stars with Exotic Phases{\LARGE \footnote{Based on the talk
in the International Conference "Baryons13" held at Glasgow in June 2013. 
%and the workshop held at Kyoto in August 2013
}} \vspace*{2mm}
}
\end{center}
%\vspace{5mm}

\begin{center}

\vspace*{2mm}
%\vspace*{5mm}
{\large Motoi Tachibana${}^{\dagger}$\footnote[2]{\tt motoi@cc.saga-u.ac.jp} and
Marco Ruggieri${}^{\ddagger}$\footnote[3]{\tt marco.ruggieri@lns.infn.it}
}\\

%\vspace*{5mm}
\vspace*{2mm}
{${}^{\dagger}$Department of Physics, Saga University, Saga 840-8502, Japan\\}
{${}^{\ddagger}$Department of Physics and Astronomy, University of Catania, 
} \\
{%\large 
Via S. Sofia 64, I-95125 Catania, Italy\\}
%\vspace*{2mm}
%{${}^{\dagger}$Department of Physics, Saga University, Saga 840-8502, Japan\\}
%\vspace*{2mm}
%{\em ${}^c$ Institut f\"ur Theoretische Physik, ETH Z\"urich, \\ 
%CH-8093 Z\"urich, Switzerland}
%\vspace*{2mm}

\vspace*{3mm}
\end{center}

\begin{center}
{\large Abstract}
\end{center}
In this short paper, we argue the issue on dark matter capture in neutron stars.
After summarizing the whole scenario and the introduction of previous studies along this line, 
we propose some potentially important effects due to the appearance of exotic phases
such as neutron superfluidity, meson condensation and quark superconducitivity.
%which have been seriously taken into account before.
Those effects might be sizable and alter the previous results.
\noindent

\vfill
\begin{flushleft}

\end{flushleft}
\end{titlepage}
\newpage

\vspace{1cm}

%\section{Introduction}

One of the modern physics perspectives is unity of matters and universe, as shown by UROBOROS.
Study of the early universe is such an example, where particle physics and cosmology are mutually
intertwined. In this short paper, we would like to present a connection between
astrophysics and particle physics as another example.

Dark matter (DM) was originally proposed by Zwicky as "missing mass" in 1933 \cite{DM}. Since then,
there are enormous indirect evidences of its existence such as the galaxy rotation curve
and the cosmic microwave background (CMB). So many people have no doubt for the existence
of DM, while the properties are little known and mysterious. One candidate for the DM is weakly-interacting massive
particle (WIMP), which interacts with nucleons through the weak interaction.
DM is one of the most interesting subjects in modern physics.

On the other hand, neutron star (NS) was proposed by Baade and Zwicky in 1934,
as a remnant after the supernova explosion \cite{NS}. Landau called the star "a gigantic nucleus"
since it is literally made by neutrons \cite{Landau}. From the properties such as density,
magnetic field and rotation period,
we see that NS provides a good market selling ultimate environments, which leads us to
the concept of dense nuclear matter and even denser object, i.e., deconfined quark matter.

In spite of many indirect evidences, we need direct measurements of DM to comprehend its properties.
To this end, there are on-ground experiments, such as XENON and CDMS \cite{DIRECT}. 
Those experiments try to constrain the WIMP DM-nucleon cross section as well as the WIMP DM mass. 

Here is an interesting observation. If dark matter really exists, it should be everywhere, 
even around neutron stars.  When a DM comes inside NS, it interacts with neutrons.
Since the cross section $\sigma_{N\chi}$ is given by the mean free path as well as
the number density of medium, one can roughly estimate $\sigma_{N\chi}$ in the case of typical
neutron star mass and radius, $M_{NS}=1.4M_{SUN}$ and $R=10$km, where $M_{SUN}$ is the solar mass. 
The result is  $\sigma_{N\chi} \approx 5\times10^{-46} \ $cm$^2$, which is the way below the limit
through the direct measurements.
From this naive observation, we are encouraged to pursue the connection between
dark matter and neutron stars in more detail. 

There have already been various studies along this line \cite{review}. Among them, 
we consider here the issue on dark matter capture in neutron stars and propose some
new ideas from the viewpoint of many-body physics. Note here that the idea of DM capture by
the stellar objects itself is not so new. For instance, people have considered
the DM capture by Sun as a solution to the solar neutrino puzzle \cite{Press-Spergel} and applied
the idea into the case of NS \cite{Goldman-Nussinov}.

In considering this issue, there are three important stages. I. Capture, II. Thermalization
and III. Black-hole formation. Here we summarize the previous studies, especially
based on \cite{MYZ}.

The stage I is described by the following equation:
\beq
\frac{dN_{\chi}}{dt}=C_{N\chi}+C_{\chi\chi}N_{\chi}-C_{\chi a}N_{\chi}^2, \nonumber
\label{capture}
\eeq 
where $N_{\chi}$ is the number of captured DMs. $C_{N\chi}, C_{\chi\chi}$
and $C_{\chi a}$ are the DM-neutron capture rate, the DM self-capture rate
and the DM pair annihilation rate, respectively. 

In the stage II, DM loses its energy through the collision with neutrons, and
then gets into the thermal equilibrium. The process is described by the following equation:
\beq
\frac{dE}{dt}=-\xi n_B\sigma_{N\chi}v\delta E, \nonumber
\label{thermalization}
\eeq 
where $E$ is the energy of DM, $\delta E$ the energy loss per one collision,
$\sigma_{N\chi}$ the DM-neutron cross section, $n_B$ the baryon number density,
and $v$ the DM velocity, respectively. Remark here on the parameter $\xi$, which
is called the capture efficiency factor. In neutron star, neutrons are highly degenerate
so that there is the Pauli-blocking effect. $\xi$ exactly parametrizes how much the system
is degenerate.

Lastly in the stage III, thermalized DMs drift into the core of NS and form the
isothermal sphere, whose radius $r_{th}$ is fixed by the balance between
kinetic and gravitational potential energy. As time goes by, the number of
captured DMs increases. If the DM density gets larger than
the baryon density $\rho_B$ within the radius $r_{th}$, namely,
\beq
\frac{3N_{\chi}m_{\chi}}{4\pi r_{th}^3} \geq \rho_B \nonumber
\label{self-gravitation}
\eeq 
then DMs become self-gravitating.
This is the on-set of gravitational collapse and black-hole formation. From the above inequality,
one can estimate the critical number of DMs, $N_{self}$, beyond which the host 
neutron star can be destructed. This is an essential point because 
relatively old neutron stars with the age around $10^{10}$ years have been found observationally. 
This means that the following inequality should hold:
\beq
N_{\chi} \leq N_{self}, \nonumber
\label{destruction}
\eeq 
which gives the constraint between the DM parameters, $m_{\chi}$
and $\sigma_{N\chi}$.

In \cite{MYZ}, as concrete examples, the astrophysical observations
such as the pulsar B1620-26 and the nearby pulsars
J0437-4715 and J2124-3358 have been investigated. Consequently 
the constraints on DM parameters 
%$\sigma_{N\chi}$ and $m_{\chi}$
obtained there are well below the limit from the direct measurements.

The above is a brief summary of the scenario for the DM capture in NS.
So far, people have been studying the issue mainly
from particle physics side. As was mentioned before, however, NS is literally a star
made by highly degenerate neutrons so that many-body
effects, which have been neglected in the preceding works\footnote{The only 
exception is the treatment performed by Bertoni et al. \cite{BNR}.},
should be important and may alter the previous results.

According to the many-body calculations, depending on temperature and 
chemical potential of the system, the existence of 
exotic phases such as neutron superfluidity, proton superconductivity,
meson condensation and quark superconductivity/superfluidity have been predicted.
Those phases can be characterized by the energy gap of quasiparticles
and the appearance of some new low-energy degrees of freedom
associated with symmetry breaking.\\
\\
Here are examples to show what could happen if the exotic phases appear in NS \cite{MOTOI}:\\
\\
(1) Neutron superfluid phase\\
In this phase, the superfluid (SF) phonon mode appears as a new degrees of freedom
at low energy. The dominant process which affects the capture rate
and thermalizaion of DM stems from the scattering between DM and SF phonons.
This can be described by the effective Lagrangian, for instance, such as that in \cite{JR}.\\
\\
(2) Color-flavor locked (CFL) phase\\
This phase is expected to occur at asymptotically high baryon density.
In the phase, all the quarks gain the energy gap $\Delta_{CFL}$ and
the low-energy degrees of freedom are the Nambu-Goldstone modes
associated with $U(1)_B$ symmetry breaking as well as chiral symmetry breaking.
$\Delta_{CFL}$ affects the structure of the Fermi surface, which
leads us to the modification of the capture efficiency factor $\xi$.
Since $\Delta_{CFL}$ is estimated around several tenth of MeV,
compared with the Fermi momentum $p_F$,
the effect could be sizable. This point will be addressed in the future 
%\cite{RT}
.\\

In summary, stellar constraints on DM properties, together with
the direct measurements and the collider experiments, are important.
In this manuscript, we specially considered the issue on the DM capture
in NS.
%which may give severe constraints on the DM properties.
Since the existence of exotic phases due to the many-body
effects have not been so much cared in the previous studies,
we proposed here what could happen if the exotic phases are taken into account. 
More detailed study will be the future issue.

\section*{Acknowledgements}
%We would like to thank  for fruitful comments and discussions. 
The authors thank the Yukawa Institute for Theoretical Physics, Kyoto University. Discussions during the YITP workshop YITP-T-13-05 on ''New Frontiers in QCD" were useful to complete this work. 
The work of M.T. is supported in part by the JSPS Grant-in-Aid for Scientific Research, Grant No. 24540280. 

%%%%%%%%%%%%%%%%%%% APPENDIX %%%%%%%%%%%%%%%%%%%%%%%

%\noindent{\bf\Large Appendix}
%\appendix
%\def\thesection{Appendix \Alph{section}}

%%%%%%%%%%%%%%%%%%%%%%%%%%%%%%%%%%%%%%%%%%%%%%%%%%
%%%%%%%% References (without bibtex) %%%%%%%%%%%%%
%%%%%%%%%%%%%%%%%%%%%%%%%%%%%%%%%%%%%%%%%%%%%%%%%%

%%%%%%%%%%%%%%%%%%%%%%%%%%%%%%%%%%%%%%%%%%%%%%%%%%
%%%%%%%%%%% References bibtex%%%%%%%%%%%%%%%%%%%%%
%%%%%%%%%%%%%%%%%%%%%%%%%%%%%%%%%%%%%%%%%%%%%%%%%%
 % \bibliographystyle{JHEP}
 % \bibliography{ref-averaged-inst-note}

\newpage

\end{document}